\documentclass{aa}

\usepackage{graphicx}
\usepackage{txfonts}
\RequirePackage{rotating}
\RequirePackage{wrapfig}
\usepackage{graphicx}   
\usepackage{amsmath}    
\usepackage{amssymb}    
\usepackage[graphicx]{realboxes}
\usepackage[T1]{fontenc}
\usepackage{ae,aecompl}
\usepackage{hyperref}
\usepackage{pdflscape}
\usepackage{newtxtext,newtxmath}
\usepackage{wrapfig}
\usepackage{subfig}
\usepackage{xcolor}

\begin{document}

\def\simlt{\mathrel{\rlap{\lower 3pt\hbox{$\sim$}}\raise 2.0pt\hbox{$<$}}}
\def\simgt{\mathrel{\rlap{\lower 3pt\hbox{$\sim$}} \raise 2.0pt\hbox{$>$}}}

   \title{{Galaxy populations} and redshift dependence of the correlation between infrared and radio luminosity}

   \subtitle{}

   \author{G. De Zotti
          \inst{1}
          \and
          M. Bonato\inst{2}
           \and
M. Giulietti\inst{3}
\and
M. Massardi\inst{2,4}
\and
M. Negrello\inst{5}
\and
H. S. B. Algera\inst{7,8}
\and
J. Delhaize\inst{9}
          }

   \institute{INAF, Osservatorio Astronomico di Padova, Vicolo Osservatorio 5, I-35122 Padova, Italy\\
              \email{gianfranco.dezotti@inaf.it}
         \and
INAF, Istituto di Radioastronomia - Italian ARC, Via Piero Gobetti 101, I-40129 Bologna, Italy
\and
INAF, Istituto di Radioastronomia, Via Piero Gobetti 101, 40129 Bologna, Italy
\and
SISSA, Via Bonomea 265, I-34136 Trieste, Italy
\and
School of Physics and Astronomy, Cardiff University, The Parade, CF24 3AA, UK
\and
Hiroshima Astrophysical Science Center, Hiroshima University, 1-3-1 Kagamiyama, Higashi-Hiroshima, Hiroshima 739-8526, Japan
\and
National Astronomical Observatory of Japan, 2-21-1, Osawa, Mitaka, Tokyo, Japan
\and
Department of Astronomy, University of Cape Town, 7701 Rondebosch, Cape Town, South Africa
            }

   \date{...}


  \abstract{We argue that the difference in infrared-to-radio luminosity ratio between local and high-redshift star-forming galaxies reflects {the alternative physical conditions} ---including magnetic field configurations--- of the dominant population of star-forming galaxies in different redshift ranges. We define three galactic types, based on our reference model, with reference to ages of stellar populations. ``Normal'' late-type galaxies dominate the star formation in the nearby  Universe; ``starburst'' galaxies take over at higher redshifts, up to $z\sim 1.5$; while ``protospheroidal'' galaxies dominate at high redshift. A reanalysis of data from the COSMOS field combined with literature results shows that, for each population, the data are consistent with an almost redshift-independent mean value of the parameter $q_{\rm IR}$, which quantifies the infrared--radio correlation. However, we find a hint of an upturn of the mean $q_{\rm IR}$ at $z\simgt 3.5$ consistent with the predicted dimming of synchrotron emission due to cooling of relativistic electrons by inverse Compton scattering off the cosmic microwave background. The typical stellar masses increase from normal, to starburst, and to protospheroidal galaxies, accounting for the reported dependence of the mean $q_{\rm IR}$ on stellar mass. Higher values of $q_{\rm IR}$ found for high-$z$ strongly lensed dusty galaxies selected at $500\,\mu$m might be explained by differential magnification.}
  \keywords{galaxies: general -- radio continuum: galaxies -- infrared: galaxies -- galaxies: starburst --galaxies: starburst -- galaxies: high-redshift
               }

   \maketitle

\section{Introduction}\label{sect:introduction}

The surprisingly tight correlation between the far-infrared {(FIR; rest-frame 42.5--$122.5\,\mu$m)} emission of local star-forming galaxies (SFGs) and their {monochromatic} nonthermal radio emission \citep[FIRRC;][]{DickeySalpeter1984, deJong1985, Helou1985, Gavazzi1986, Condon1992, Yun2001} has been extensively investigated. The FIRRC qualifies radio observations of galaxies as star formation diagnostics unaffected by dust extinction, and opens the possibility to exploit the ongoing and forthcoming large-area and deep radio surveys ---notably those by the Square Kilometer Array (SKA) and its precursors--- to trace the cosmic star-formation history in unprecedented detail up to very high redshifts \citep{Haarsma2000, Jarvis2015, Mancuso2015a, Mancuso2015b, DeZotti2019, Cochrane2023, Ocran2023}. However, this utilization is complicated by our persistent lack of understanding regarding its redshift dependence.

The correlation is conventionally described in terms of the parameter $q_{\rm IR}$ , which is defined as \citep{Bell2003, Ivison2010a, Ivison2010b, Sargent2010}:
\begin{equation}\label{eq:q_ir}
q_{\rm IR}=\log\left(\frac{L_{\rm IR}[\hbox{W}]/3.75\times 10^{12}}{L_{1.4\,\rm GHz} [\hbox{W}\,\hbox{Hz}^{-1}]}\right),
\end{equation}
where $L_{\rm IR}$ is the total IR luminosity {(rest-frame 8--$1000\,\mu$m)}.

Extensive studies of samples of nearby star-forming galaxies, with $L_{\rm IR}$ derived from the InfraRed Astronomical Satellite (IRAS) data, have yielded a mean value of $\bar{q}_{\rm IR}\simeq 2.6$ \citep{Condon1992, Yun2001, Bell2003}\footnote{The values of $q_{\rm FIR}$, referring to FIR luminosities, have been scaled assuming $L_{\rm IR}/L_{\rm FIR}\simeq 2$ \citep{Bell2003}.}. A slightly lower value ($\bar{q}_{\rm IR}\simeq 2.54$) was reported by \citet{Molnar2021} for a large sample of IR- and radio-detected galaxies at $z < 0.2$, taking into account also Wide-Field Infrared Survey Explorer (WISE) and \textit{Herschel} photometry (the latter being available for a minor fraction of galaxies). A value of $\bar{q}_{\rm IR}$ similar to that by \citet{Molnar2021} was found by \citet{Tisanic2022} for a \textit{Herschel}-selected sample at $z < 0.1$.

\textit{Herschel} surveys have allowed investigation of the FIRRC up to $z\sim 5$. The reported high-$z$ values of $\bar{q}_{\rm IR}$ are generally lower than in the nearby Universe, implying higher radio-to-IR luminosity ratios\footnote{See, however, \citet{Sargent2010} who, using the 1.4\,GHz VLA map and the \textit{Spitzer} coverage of the COSMOS field, did not find any change of $\bar{q}_{\rm IR}$ out to $z\sim 2$ for ultra luminous infrared galaxies (ULIRGs) and IR-luminous galaxies.}. The redshift dependence of $\bar{q}_{\rm IR}$ has been expressed in the form $\bar{q}_{\rm IR}\propto (1+z)^{-\beta}$ with $\beta$ ranging from $\simeq 0.12$ to $\simeq 0.20$ \citep{Magnelli2015, Basu2015, CalistroRivera2017, Delhaize2017, Ocran2020, Sinha2022}.

The physical origin of the redshift dependence of $\bar{q}_{\rm IR}$ is unclear. It has been suggested that it may be due to selection biases. Several studies have reported evidence of an increase in the radio-to-IR luminosity ratio with increasing stellar mass \citep{Gurkan2018, Delvecchio2021, Smith2021, McCheyne2022}. This translates to a decrease in $\bar{q}_{\rm IR}$ for flux-limited surveys, which are biased toward increasingly massive galaxies at higher and higher redshifts. However, there are significant differences in the inferred strength of the stellar mass dependence {related, for instance, to sample selection, active galactic nuclei (AGN) subtraction, and IR luminosity derivation}; but no dependence was seen by \citet{Pannella2015}.

A redshift-dependent bias on $\bar{q}_{\rm IR}$ may also be the consequence of a slightly superlinear relation between $L_{\rm radio}$  and $L_{\rm IR}$ of the form $L_{\rm radio}\propto L_{\rm IR}^b$ with $b\simeq 1.11$ \citep{Basu2015, Molnar2021}. The decrease in $\bar{q}_{\rm IR}$ with increasing $z$ may follow from the preferential sampling of increasingly luminous sources.

In this paper, we discuss an alternative possibility based on the consideration that different galaxy populations dominate the cosmic star-formation activity in different redshift ranges. {In Sect.\,\ref{sect:scenario} we describe our reference scenario and motivate its choice. In} Sect.\,\ref{sect:samples} we reanalyze the ``jointly-selected'' sample by \citet{Delhaize2017}, a combination of radio- and infrared-selected samples. We also analyze the sample of \textit{Herschel}-selected candidate strongly lensed submillimeter galaxies (SMGs) by \citet{Negrello2017}, and include the results on the FIRRC of high-$z$ SMGs by \citet{Algera2020}.

{The \citet{Delhaize2017} sample is particularly important for our purposes because it minimizes the selection bias and covers a broad redshift range  with good statistics; hence it is very well suited to investigation of the redshift dependence of $\bar{q}_{\rm IR}$. The samples by \citet{Negrello2017} and \citet{Algera2020} add information on $\bar{q}_{\rm IR}$ at high redshifts, where the uncertainties affecting the \citet{Delhaize2017} sample are higher. One further point of interest regarding the \citet{Negrello2017} sample is that it is made of confirmed and candidate strongly lensed galaxies. Strong lensing offers the possibility to study faint galaxies that would be otherwise difficult to observe \citep{Giulietti2022}. On the other hand, a bias could be introduced by differential magnification, an issue that merits investigation.} In Sect.\,\ref{sect:conclusions} we discuss our results and present our conclusions.

We adopt a flat cosmology with $H_0=70\,\hbox{km}\,\hbox{s}^{-1}\,\hbox{Mpc}^{-1}$, $\Omega_{\rm M}=0.3,$ and $\Omega_\Lambda=0.7$.

\begin{figure}
        \includegraphics[width=\linewidth]{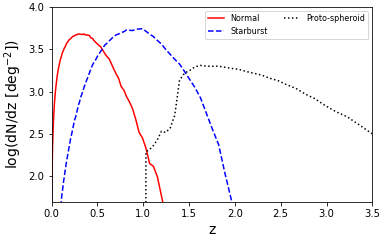}
  \caption{Redshift distributions of ``normal'' late-type, ``starburst'', and ``protospheroidal'' galaxies predicted by the \citet{Bonato2017} model for $S_{1.4\,\rm GHz} > 10\,\mu$Jy, corresponding to approximately five times the rms fluctuations of the MIGHTEE survey of the COSMOS field \citep[][{see Sect.\ref{sect:samples}}]{Heywood2022}. {See text for our definition of the three galaxy populations.}
}
  \label{fig:zdistr}
\end{figure}

\section{{Evolutionary scenario}}\label{sect:scenario}

{Our physics-based reference model is the one developed by \citet{Cai2013}, building on the work of \citet{Granato2004} and \citet{Lapi2006, Lapi2011}. In spite of its age, the model is still highly competitive \citep[see, e.g.,][]{Chen2022, Ward2022, Vargas2023, RowanRobinson2024}. The validity of the underlying evolutionary scenario is further supported by the data-driven, model-independent approach by \citet{Boco2023}.}

Analyses of the global galaxy stellar populations of nearby galaxies \citep{Thomas2010, Bernardi2010, Lu2023} have highlighted a clearly bimodal age distribution implying that early-type galaxies had high star formation rates (SFRs) at early times ($\log(\hbox{age}/\hbox{yr})\simgt 9.5$ corresponding to $z\simgt 1.5$) and were quenched afterwards {\citep{vanDokkum2015}. During their active star-forming phase, these galaxies show up as SMGs. Several studies have indeed pointed out that the properties of high-$z$ SMGs are consistent with them being the progenitors of present-day massive elliptical galaxies \citep{Lilly1999, Swinbank2006, Hainline2011, Toft2014, Simpson2014, Simpson2017, Amvrosiadis2023, Liao2024, Birkin2024}. In our reference scenario, as their stellar mass grows and their dust obscuration decreases, high-$z$ SMGs evolve to the main sequence \citep{Mancuso2016}, that is, to the population of high-$z$ massive star-forming galaxies detected in optical/near-infrared(NIR) spectroscopic surveys, with lower dust-to-stellar-mass ratios and substantially lower specific SFR compared to SMGs. In this phase, high-$z$ SMGs frequently do not show a spheroidal morphology. Most are gas-rich rotation-dominated disks, albeit dynamically hotter (i.e., with a substantially lower ratio of rotational to random motions) and with a larger fraction of irregular morphologies \citep{Swinbank2017, Turner2017, Wisnioski2019, ForsterSchreiberWuyts2020, Kartaltepe2023, Birkin2024} than spiral galaxies in the local Universe \citep{Lelli2016}, with few exceptions \citep{Rizzo2020}. The transition to present-time massive early-type galaxies implies kinematic and  morphological evolution. This population is referred to here as ``protospheroidal galaxies'' or ``protospheroids''.}

On the contrary, {present-day disk galaxies with high ratios of rotational to random velocity, as in the Milky Way, never achieved very high SFRs but maintained a relatively low, slowly evolving star formation activity up to their formation epoch, which is supported by detailed evolutionary models \citep[e.g.,][]{Snaith2015, Valentini2019, Giammaria2021}; their SFRs are generally of a few to several $M_\odot\,\hbox{yr}^{-1}$. We refer to this population as ``normal late-type galaxies''. Based on the cited models, in the absence of morphological information, we assign to this population galaxies with an SFR of lower than $\hbox{SFR}_{\rm lim} = 10\, M_\odot\,\hbox{yr}^{-1}$.}

We note that {$\hbox{SFR}_{\rm lim}$ or $\log(L_{\rm IR}/L_\odot)=11$ are conventionally adopted as the lower SFR/$L_{\rm IR}$ limits for luminous infrared galaxies \citep[LIRGs;][]{Elbaz2002, Lagache2005}. Low- to moderate-redshift ($z\simlt 1-1.5$) LIRGs and more intensely star-forming galaxies (ultraluminous or hyperluminous infrared galaxies, ULIRGs or HyLIRGs; $\log(L_{\rm IR}/L_\odot)\simgt 12$) cover the same IR luminosity ranges as protospheroids, but their astrophysical properties are substantially different \citep[and references therein]{Rujopakarn2011}. Differences are found in their spectral energy distributions (SEDs), metallicity, and structure. The sizes of the star-forming regions of protospheroids are more extended than those of low/moderate-$z$ (U)LIRGs of similar IR luminosity \citep{Younger2010, Iono2016, Mitsuhashi2023}. Protospheroids generally have much higher velocity dispersion/rotation velocity ratios \citep{Genzel2008, Law2009}.  Also, the fraction of protospheroids showing merger morphology is lower than that of low-$z$ ULIRGs \citep{Kartaltepe2010}. While protospheroids are in the process of forming most of their stars, low/moderate-$z$ (U)LIRGs possess an old stellar population and can be interpreted as late-type galaxies undergoing a strong burst of star formation  triggered by interactions and mergers or early-type galaxies ``rejuvenated'' by wet mergers.  }

{We note that definitions of normal and starburst galaxies that can be found in the literature differ from ours. \citet{Elbaz2002} and \citet{Lagache2005} refer to star-forming galaxies with $(\hbox{SFR}/M_\odot\,\hbox{yr}^{-1})\simlt 1$ and $1\simlt (\hbox{SFR}/M_\odot\,\hbox{yr}^{-1})\simlt 10$ as normal and starburst,  respectively. More recently, it has become common to distinguish galaxies forming stars in normal main sequence mode from starbursts showing higher specific SFRs \citep{Elbaz2011, Rodighiero2011, Pannella2015, Popesso2023}.  Main sequence  galaxies at fixed stellar mass had much higher SFRs at high $z$ than they have today. In our scenario, the main sequence is not an evolutionary sequence, which is consistent with the analysis by \citet{Mancuso2016}.}



The statistical properties of the three populations (protospheroidal, starburst, and {normal late-type} galaxies) have been investigated over a broad range of wavelengths, including the radio by \citet{Cai2013}, \citet{Mancuso2015b}, and \citet{Bonato2017}. {The latter authors converted the redshift-dependent SFR functions for the three populations yielded by the \citet{Cai2013} model to evolving radio luminosity functions using two relationships between SFR and radio luminosity taken from the literature. We have adopted the so-called ``nonlinear'' relation (synchrotron luminosity proportional to $\hbox{SFR}^{1.1}$). The model was shown to accurately reproduce the observed radio luminosity functions up to the highest redshifts ($z\simeq 5.1$) at which they were observationally determined.}

The redshift distributions {predicted by the \citet{Bonato2017} model} for $S_{1.4\,\rm GHz} > 10\,\mu$Jy, which is approximately five times the root mean square (rms) fluctuations of the MIGHTEE survey of the COSMOS field \citep[][{see Sect.\ref{sect:samples}}]{Heywood2022}, are shown in Fig.\,\ref{fig:zdistr}. Normal late-type galaxies dominate at $z<0.5$, starburst galaxies in the range $0.5<z<1.5$, and protospheroids  at larger redshifts.

Differences in the FIRRCs between {different galaxy populations are} expected \citep[see][for a discussion of this issue]{Algera2020}. Spheroidal galaxies have, on average, higher stellar masses \citep{Molnar2018}, and therefore higher radio to IR luminosity ratios are expected based on the evidence mentioned above of an increase in this ratio with stellar mass.  Also, the high SFRs of high-$z$ protospheroidal galaxies imply high
rates of cosmic-ray production via supernova explosions. \citet{Lacki2010} argued that these galaxies are calorimetric and possess enhanced magnetic fields, meaning that synchrotron energy losses dominate over the other cosmic-ray energy loss mechanisms (inverse Compton, bremsstrahlung, and ionization), {leading to enhanced radio-to-IR-luminosity ratios or lower $\bar{q}_{\rm IR}$}. {A lower $\bar{q}_{\rm IR}$ of low-$z$ starburst compared to normal galaxies may again} be related to the magnetic field strengths, which are substantially larger for starbursts \citep[e.g.,][]{Yoast-Hull2016}.
The above reasoning leads us to question whether or not the observed redshift dependence of $\bar{q}_{\rm IR}$ could be due to the fact that we sample different galaxy populations in different redshift intervals, while the radio--IR correlation of each population is redshift-independent in the $z$ range probed so far.

\section{Galaxy samples}\label{sect:samples}

\subsection{The \citet{Delhaize2017} sample}

\begin{figure}
        \includegraphics[width=\linewidth]{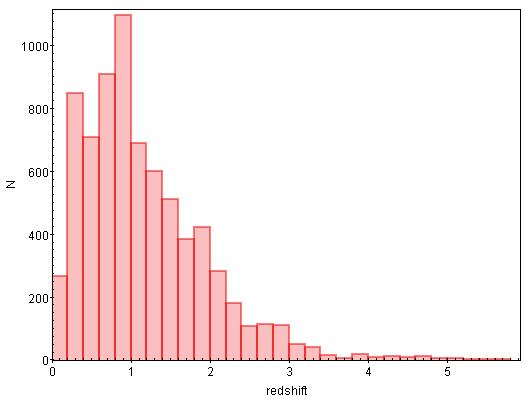}
  \caption{Redshift distribution for the Delhaize/MIGHTEE sample (see text).}
  \label{fig:Delhaize_zdistr}
\end{figure}

\begin{table}
\begin{center}
\caption{Median and mean values of $q_{\rm IR}$ as a function of redshift for the Delhaize/MIGHTEE sample. First column: redshift interval; second column: total number of sources in the interval; third column: number of lower limits to $q_{\rm IR}$; fourth, fifth, and sixth columns: values of $q_{\rm IR}$ corresponding to the first, second (i.e., median), and third quartile; seventh and eight columns: mean $q_{\rm IR}$ and its error given by ASURV. The 1.4\,GHz luminosities and upper limits derived from the 3\,GHz survey have been increased by a factor of 1.51 (see text).}
\label{tab:Delhaize_qIR}
\resizebox{\columnwidth}{!}{\begin{tabular}{rrrrrrrr}
\hline
$z$ range &$N_{\rm s}$ & $N_{\rm l}$ & I quart. & Median & III quart. & Mean & Err. \\
\hline
0.00--0.10 &  54 &  24 &  2.520 & 2.646 & 2.749 & 2.693 & 0.035 \\
0.10--0.20 & 213 &  67 &  2.438 & 2.560 & 2.721 & 2.594 & 0.017 \\
0.20--0.30 & 306 &  87 &  2.391 & 2.507 & 2.659 & 2.529 & 0.013 \\
0.30--0.35 & 264 &  81 &  2.389 & 2.509 & 2.612 & 2.505 & 0.013 \\
0.35--0.40 & 277 &  88 &  2.360 & 2.498 & 2.614 & 2.478 & 0.012 \\
0.40--0.45 & 147 &  41 &  2.334 & 2.456 & 2.587 & 2.453 & 0.015 \\
0.45--0.50 & 186 &  62 &  2.334 & 2.458 & 2.583 & 2.463 & 0.018 \\
0.50--0.55 & 208 &  58 &  2.275 & 2.414 & 2.529 & 2.422 & 0.018 \\
0.55--0.60 & 165 &  52 &  2.286 & 2.406 & 2.551 & 2.421 & 0.021 \\
0.60--0.65 & 164 &  46 &  2.277 & 2.403 & 2.542 & 2.410 & 0.018 \\
0.65--0.70 & 325 & 105 &  2.298 & 2.406 & 2.535 & 2.404 & 0.012 \\
0.70--0.75 & 274 &  94 &  2.266 & 2.408 & 2.572 & 2.407 & 0.015 \\
0.75--0.80 & 145 &  45 &  2.215 & 2.367 & 2.537 & 2.377 & 0.020 \\
0.80--0.85 & 269 &  93 &  2.246 & 2.405 & 2.550 & 2.417 & 0.026 \\
0.85--0.90 & 276 & 104 &  2.229 & 2.391 & 2.546 & 2.468 & 0.030 \\
0.90--0.95 & 280 & 104 &  2.258 & 2.398 & 2.560 & 2.418 & 0.020 \\
0.95--1.00 & 269 & 101 &  2.229 & 2.377 & 2.546 & 2.418 & 0.021 \\
1.00--1.10 & 341 & 133 &  2.195 & 2.371 & 2.562 & 2.369 & 0.016 \\
1.10--1.20 & 348 & 138 &  2.177 & 2.358 & 2.530 & 2.352 & 0.017 \\
1.20--1.30 & 326 & 115 &  2.186 & 2.358 & 2.481 & 2.340 & 0.018 \\
1.30--1.40 & 274 &  96 &  2.134 & 2.326 & 2.476 & 2.305 & 0.018 \\
1.40--1.50 & 276 &  88 &  2.139 & 2.313 & 2.468 & 2.290 & 0.018 \\
1.50--1.60 & 233 &  81 &  2.078 & 2.322 & 2.541 & 2.327 & 0.034 \\
1.60--1.70 & 199 &  68 &  2.057 & 2.265 & 2.455 & 2.255 & 0.021 \\
1.70--1.80 & 187 &  73 &  2.096 & 2.286 & 2.471 & 2.284 & 0.026 \\
1.80--1.90 & 198 &  77 &  2.036 & 2.250 & 2.500 & 2.299 & 0.032 \\
1.90--2.00 & 226 &  71 &  2.040 & 2.253 & 2.426 & 2.239 & 0.022 \\
2.00--2.10 & 159 &  49 &  2.061 & 2.271 & 2.413 & 2.225 & 0.023 \\
2.10--2.20 & 124 &  43 &  2.111 & 2.269 & 2.431 & 2.338 & 0.040 \\
2.20--2.30 & 114 &  36 &  1.949 & 2.219 & 2.437 & 2.204 & 0.030 \\
2.30--2.40 &  68 &  21 &  1.929 & 2.236 & 2.496 & 2.202 & 0.040 \\
2.40--2.50 &  47 &  16 &  2.001 & 2.220 & 2.425 & 2.223 & 0.042 \\
2.50--2.60 &  63 &  21 &  1.961 & 2.184 & 2.454 & 2.217 & 0.041 \\
2.60--2.70 &  54 &  18 &  1.892 & 2.183 & 2.468 & 2.171 & 0.049 \\
2.70--2.80 &  60 &  13 &  1.998 & 2.237 & 2.456 & 2.219 & 0.039 \\
2.80--2.90 &  55 &  16 &  1.994 & 2.270 & 2.382 & 2.230 & 0.047 \\
2.90--3.00 &  55 &  13 &  2.012 & 2.215 & 2.377 & 2.183 & 0.036 \\
3.00--3.10 &  31 &   7 &  2.055 & 2.168 & 2.323 & 2.193 & 0.050 \\
3.10--3.30 &  40 &   6 &  1.902 & 2.172 & 2.414 & 2.135 & 0.051 \\
3.30--3.50 &  25 &   3 &  1.797 & 2.109 & 2.376 & 2.122 & 0.067 \\
3.50--4.00 &  29 &   7 &  1.954 & 2.354 & 2.592 & 2.315 & 0.065 \\
4.00--4.50 &  24 &   5 &  2.043 & 2.399 & 2.557 & 2.346 & 0.065 \\
4.50--5.00 &  19 &   6 &  2.333 & 2.541 & 2.629 & 2.474 & 0.088 \\
5.00--5.70 &  10 &   1 &  1.917 & 2.049 & 2.545 & 2.245 & 0.126 \\
\hline
\end{tabular} }
\end{center}
\end{table}

\begin{table}
\begin{center}
\caption{Median and mean values of $q_{\rm IR}$ as a function of redshift for the {``starburst plus protospheroid''} sample ($\hbox{SFR}> 10\,M_\odot\,\hbox{yr}^{-1}$). The 1.4\,GHz luminosities and upper limits derived from the 3\,GHz survey have been scaled up by a factor 1.51 (see text). The columns have the same meaning as in Table\,\protect\ref{tab:Delhaize_qIR}. }
\label{tab:starburst_qIR}
\resizebox{\columnwidth}{!}{\begin{tabular}{rrrrrrrr}
\hline
$z$ range &$N_{\rm s}$ & $N_{\rm l}$ & I quart. & Median & III quart. & Mean & Err. \\
\hline
0.15--0.35 &  38 &   0 & 2.343 & 2.478 & 2.580 & 2.474 & 0.021 \\
0.35--0.45 &  70 &   2 & 2.367 & 2.448 & 2.542 & 2.456 & 0.016 \\
0.45--0.55 & 157 &  24 & 2.330 & 2.442 & 2.539 & 2.451 & 0.017 \\
0.55--0.65 & 163 &  30 & 2.334 & 2.423 & 2.547 & 2.442 & 0.015 \\
0.65--0.75 & 439 & 132 & 2.317 & 2.432 & 2.582 & 2.438 & 0.009 \\
0.75--0.85 & 354 & 119 & 2.279 & 2.420 & 2.565 & 2.447 & 0.022 \\
0.85--0.95 & 499 & 193 & 2.288 & 2.419 & 2.565 & 2.464 & 0.024 \\
0.95--1.05 & 431 & 175 & 2.254 & 2.405 & 2.560 & 2.436 & 0.016 \\
1.05--1.15 & 286 & 112 & 2.195 & 2.346 & 2.554 & 2.339 & 0.015 \\
1.15--1.25 & 385 & 138 & 2.191 & 2.344 & 2.498 & 2.349 & 0.014 \\
1.25--1.35 & 252 &  93 & 2.157 & 2.366 & 2.490 & 2.359 & 0.024 \\
1.35--1.45 & 291 & 100 & 2.167 & 2.313 & 2.486 & 2.320 & 0.018 \\
1.45--1.55 & 245 &  86 & 2.105 & 2.325 & 2.498 & 2.291 & 0.021 \\
1.55--1.65 & 231 &  77 & 2.078 & 2.291 & 2.525 & 2.309 & 0.029 \\
1.65--1.75 & 160 &  64 & 2.045 & 2.308 & 2.442 & 2.242 & 0.024 \\
1.75--1.85 & 227 &  89 & 2.092 & 2.283 & 2.497 & 2.313 & 0.030 \\
1.85--1.95 & 174 &  57 & 2.008 & 2.232 & 2.438 & 2.221 & 0.025 \\
1.95--2.05 & 208 &  68 & 2.056 & 2.272 & 2.435 & 2.253 & 0.021 \\
2.05--2.15 & 143 &  45 & 2.103 & 2.278 & 2.429 & 2.290 & 0.035 \\
2.15--2.25 & 118 &  37 & 1.994 & 2.199 & 2.423 & 2.219 & 0.031 \\
2.25--2.35 &  92 &  27 & 1.965 & 2.218 & 2.427 & 2.193 & 0.032 \\
2.35--2.45 &  52 &  15 & 1.998 & 2.214 & 2.446 & 2.227 & 0.042 \\
2.45--2.55 &  49 &  18 & 1.947 & 2.170 & 2.457 & 2.178 & 0.040 \\
2.55--2.65 &  66 &  24 & 1.964 & 2.272 & 2.527 & 2.250 & 0.044 \\
2.65--2.75 &  61 &  14 & 1.992 & 2.171 & 2.372 & 2.169 & 0.038 \\
2.75--2.85 &  52 &  14 & 1.916 & 2.244 & 2.432 & 2.201 & 0.047 \\
2.85--2.95 &  57 &  17 & 2.078 & 2.274 & 2.376 & 2.247 & 0.042 \\
2.95--3.05 &  43 &   9 & 1.955 & 2.209 & 2.359 & 2.172 & 0.043 \\
3.05--3.25 &  43 &   7 & 2.039 & 2.169 & 2.360 & 2.170 & 0.043 \\
3.25--3.55 &  41 &   7 & 1.797 & 2.107 & 2.400 & 2.120 & 0.057 \\
3.55--4.05 &  29 &   7 & 2.089 & 2.365 & 2.599 & 2.327 & 0.067 \\
4.05--4.55 &  20 &   2 & 2.141 & 2.396 & 2.558 & 2.371 & 0.057 \\
4.55--5.65 &  27 &   7 & 2.041 & 2.485 & 2.667 & 2.395 & 0.079 \\
\hline
\end{tabular} }
\end{center}
\end{table}

\begin{table}
\begin{center}
\caption{Median and mean values of $q_{\rm IR}$ as a function of redshift for the Negrello sample of candidate strongly lensed SFGs. The columns have the same meaning as in Table\,\protect\ref{tab:Delhaize_qIR}. }
\label{tab:Negrello_qIR}
\resizebox{\columnwidth}{!}{\begin{tabular}{rrrrrrrr}
\hline
$z$ range &$N_{\rm s}$ & $N_{\rm l}$ & I quart. & Median & III quart. & Mean & Err. \\
\hline
1.0--2.0 &  19  &   5  &  2.326  &   2.356  &  2.626   &   2.478 & 0.080  \\
2.0--2.5 &  19  &  10  &  2.288  &   2.379  &  2.810   &   2.593 & 0.097 \\
2.5--3.0 &  18  &  10  &  2.168  &   2.461  &  2.766   &   2.478 & 0.117 \\
3.0--4.0 &  20  &   9  &  2.231  &   2.507  &  2.735   &   2.574 & 0.102 \\
\hline
\end{tabular} }
\end{center}
\end{table}



A particularly extensive and accurate study of the FIRRC over a broad redshift range was carried out by \citet{Delhaize2017}. These authors exploited the data from the 3\,GHz Cosmic Evolution Survey (COSMOS) Large Project Very Large Array {VLA} survey \citep{Smolcic2017} covering the entire COSMOS field ($2\,\hbox{deg}^2$) to average rms fluctuations of $2.3\,\mu\hbox{Jy}\,\hbox{beam}^{-1}$.  The 3\,GHz flux densities were converted to 1.4\,GHz ---the frequency commonly used to study the radio--IR correlation--- using either the available 1.4\,GHz measurements or a spectral index of $\alpha =-0.7$ ($S_\nu \propto \nu^\alpha$). The radio-detected sample was cross-matched with the COSMOS2015 multiband catalog \citep{Laigle2016} to obtain their photometry-matched radio-selected sample comprising 7,729 sources. An IR-selected sample was built by means of a prior-based source extraction from \textit{Herschel}  Photodetector Array Camera and Spectrometer (PACS) and Spectral and Photometric Imaging Receiver (SPIRE) maps of the COSMOS field. \citet{Delhaize2017} found 8,458 IR-detected objects with COSMOS2015 counterparts. {The authors removed objects showing: a NIR/mid-infrared(MIR) spectrum indicative of AGN emission; $0.5-8\,$keV X-ray luminosity $L_{\rm X} > 10^{42}\,\hbox{erg}\,\hbox{s}^{-1}$; or a significant AGN component based on the MAGPHYS \citep[Multi-wavelength Analysis of Galaxy Physical Properties;][]{daCunha2008} SED fitting, as well as objects without evidence of appreciable star-formation activity and objects with properties suggestive of low-to-moderate luminosity AGN. The remaining sample, that is, the union of the radio- and IR-selected sources,} comprises 9,575 galaxies whose radio {and IR emissions are} expected to originate from star formation (jointly selected SFG sample). Only 37\% of these sources are detected both in the radio and in the IR. The median redshift is 1.02 but  the distribution extends up to $z> 4$. 

We take the above jointly selected SFG sample of  \citet{Delhaize2017} as our parent sample, and update the analysis by taking into account new radio data. \citet{Heywood2022} released the early science continuum data products of the MIGHTEE survey using the South African MeerKAT telescope at a central frequency of 1284 MHz, with rms {thermal} fluctuations of approximately $2\,\mu\hbox{Jy}\,\hbox{beam}^{-1}$. The release includes a catalog covering $1.62\,\hbox{deg}^2$ in the COSMOS field. {The angular resolution of 8.6\,arcsec largely avoids the risk of underestimating the flux densities due to the resolution bias. On the other hand, the survey becomes confusion-limited with
$1\,\sigma$ noise levels of $\sim 4.5\,\mu\hbox{Jy}\,\hbox{beam}^{-1}$ \citep{Heywood2022}.} Our further analysis is limited to the MIGHTEE area.

The area is centered at $\hbox{RA}=150^\circ .1192$,  $\hbox{DEC}=2^\circ .2058$ and has a radius of $0^\circ.7181$. In this area, there are 8,129 SFGs belonging to our parent sample. After dropping the 703 sources labeled as ``radio excess'' by \citet{Delhaize2017}, we are left with 7426 objects; 3,968 of them were detected by the 3\,GHz survey and 3,460 by the MIGHTEE survey.  
There is some complementarity among the two surveys: 989 
{``no radio-excess'' \citep[according to][]{Delhaize2017}} SFGs were only detected by MIGHTEE, 1491 
only by the 3\,GHz survey. 
For the 2,475 {``no radio-excess''} SFGs detected by both surveys, 
we choose the MIGHTEE flux density, which refers to a frequency closer to the reference frequency (1.4\,GHz) commonly used to study the radio--IR correlation. We adopt the 1.4\,GHz flux densities given by \citet{Heywood2022}, which were derived assuming a spectral index of $-0.7$.

We further removed the additional 17 sources that show radio excess based on MIGHTEE flux densities; that is, sources with $q_{\rm IR}<1.55$ 
\citep[following][]{Algera2020}. This leaves 4,938 radio-detected SFGs with no radio  excess.
%
{One of the ``radio-undetected'' sources ($\hbox{RA}=149^\circ.95786$, $\hbox{DEC}=1^\circ .94692$) has a FIRST counterpart $1.8''$ away, with $S_{1.4\,\rm GHz, peak}=1.46$\,mJy. If we were to accept this identification, it would be classified as ``radio excess'' ($\log_{10}(L_{1.4\,\rm GHz}/\hbox{W}\,\hbox{Hz}^{-1})=25.72$, $q_{\rm IR}=1.0$). To be conservative, we removed it from the sample. This leaves  2470 radio-undetected galaxies.}

{The mean ratio between the 1.4\,GHz flux densities derived from the MIGHTEE and those from} the 3\,GHz survey for the SFGs detected by both surveys is 1.51 with a dispersion of 0.56; the median ratio is 1.38; there is no indication of a redshift dependence.
The offset  is qualitatively consistent with {that of} \citet{vanderVlugt2021}, who attributed the difference in the 3 GHz source counts between COSMOS-XS and VLA-COSMOS to an under-correction for the resolution bias in the latter flux densities by a factor of $\simeq 1.4$. A comparison of ``superdeblended'' MIGHTEE versus 3\,GHz flux densities is presented by \citet{An2021}, who discuss the subsequent radio spectral indices and the impact on the inferred FIRRC.

{The interpretation of the offset in terms of the resolution bias is supported by studies of the radio sizes of star-forming galaxies.} \citet{Muxlow2020} carried out high-sensitivity e-MERLIN and VLA 1.5\,GHz observations of SFGs at $z=1$--3, finding median radio-to-optical size ratios of close to unity. This result is at variance with \citet{Jimenez-Andrade2019}, who reported a median radio size of COSMOS 3\,GHz SFGs of smaller than the optical size by a factor of 1.3--2. \citet{Muxlow2020} argued that the difference may be due to different physical scales of the emission at the two frequencies.



We increased the 1.4\,GHz flux densities derived from the 3\,GHz survey  by a factor of 1.51. {We note that MIGHTEE and VLA-3\,GHz samples are roughly matched in sensitivity, which justifies scaling radio fluxes by a fixed factor without rescaling numbers. For radio-undetected sources, we adopt the} upper limits (and the corresponding lower limits to $q_{\rm IR}$) determined by \citet{Delhaize2017}, scaled up by a factor of 1.51.

The fraction of radio-detected, no-radio-excess SFGs in the sample, hereafter referred to as the Delhaize/MIGHTEE sample, is then $4938/7408=66.7\%$. 
Most of the 7408 sources (6133, i.e., $\simeq 83\%$) were detected at $\geq 5\,\sigma$ in at least one \textit{Herschel} band. Even for sources not detected by \textit{Herschel}, IR luminosities are strongly constrained by the MAGPHYS  \citep{daCunha2008, daCunha2015, Battisti2019} SED fitting of the available multifrequency data, because MAGPHYS ensures the balance of energy absorbed and re-emitted by dust. We therefore rely on the values of $L_{\rm IR}$ computed by \citet{Delhaize2017}.

{This is at variance with the approach taken by \citet{Delhaize2017}, who regarded as upper limits the values of $L_{\rm IR}$ obtained via MAGPHYS SED fitting in the absence of \textit{Herschel} measurements. We checked whether or not the values of $q_{\rm IR}$ obtained with our approach show statistically significant differences from those computed by \citet{Delhaize2017} in the case of \textit{Herschel}-undetected sources. For this test, we considered the redshift intervals $\Delta z=0.5$ up to $z=2.5$ and $\Delta z= 1$ at higher $z$, where the number of sources per unit $z$ bin is much smaller. No significant differences were found over the full redshift range: differences are much smaller than the statistical errors and do not show any systematic trend.  {As a counter-check}, we compared the values of $L_{\rm IR}$ we adopted with those derived from the SFRs ---which should be less sensitive to the lack of FIR/submillimeter data---, using the detected sources to calibrate the $L_{\rm IR}$--SFR relation. Again we did not find any statistically significant difference.}

The redshift distribution is shown in Fig.\,\ref{fig:Delhaize_zdistr}. We have split the sample into 44 redshift intervals, each with a width of $\Delta z=0.1,$ except at $z> 3.1,$ where the width has been gradually increased to avoid an excessively low number of sources per bin  (see Table~\ref{tab:Delhaize_qIR}). For each interval, we computed the differential $q_{\rm IR}$ distribution using the Astronomy SURVival Analysis (ASURV) software package Rev. 1.2 \citep{IsobeFeigelson1990, Lavalley1992}, which implements the methods presented in \citet{FeigelsonNelson1985} for univariate problems and those in \citet{Isobe1986} for bivariate problems. 

Table~\ref{tab:Delhaize_qIR} reports, for each redshift bin, the first, second (median), and third quartile values of the $q_{\rm IR}$ distribution, as well as the mean values with their errors, as estimated by ASURV. The errors are purely statistical. The true errors must include uncertainties on photometric redshifts, on the SFG/AGN classification, and on the incomplete radio and IR flux densities; they are likely to be substantially larger, but hard to quantify. {The incompleteness in the flux densities increases with redshift and therefore translates into a redshift-dependent incompleteness in luminosity that may affect the trend of $\bar{q}_{\rm IR}$ with $z$. The application of a survival analysis mitigates but might not totally eliminate the problem.}

The dominant star-forming galaxy population in the nearby Universe is made up of low-SFR normal late-type galaxies. According to our reference model (see Sect.~\ref{sect:introduction}), this population has a slow secular evolution and is quickly overtaken by the fast-evolving {starburst} population. This leads us to question whether the increase in the mean or median $q_{\rm IR}$ at low $z$ could be due to the increasing fraction of normal late-type galaxies.

To explore this possibility, we removed galaxies with an SFR estimated using the MAGPHYS SED fitting \citep[column labeled sfr\_magphys\_best in the catalog by][]{Delhaize2017} of smaller than $10\,M_\odot\,\hbox{yr}^{-1}$  from the Delhaize/MIGHTEE sample. 
We are left with 5503 sources ({starburst plus protospheroid} sample). About one-third of the dropped 1905 sources  (10\% of the whole sample) have $\hbox{SFR}<3\,M_\odot\,\hbox{yr}^{-1}$ ($\hbox{SFR}<1\,M_\odot\,\hbox{yr}^{-1}$). Thus, most of these sources may be normal disk galaxies \citep[for reference, the Milky Way has a current SFR of about $2\,M_\odot\,\hbox{yr}^{-1}$; ][]{LicquiaNewman2015}.
For low-SFR galaxies, evolved stellar populations may contribute substantially to dust heating, and therefore to the IR luminosity, resulting in anomalously high values of $q_{\rm IR}$.

The {starburst plus protospheroid} sample has no objects at $z< 0.15$. As shown by Table\,\ref{tab:starburst_qIR}, the data are consistent with an almost redshift-independent mean $q_{\rm IR}$ of up to $z\simeq 1$: {the dispersion of the mean values is 0.013, which is consistent with statistical errors; there is no detectable trend with redshift}.

At higher $z,$ we recover the results for the Delhaize/MIGHTEE sample. There is a transition region where the mean  $q_{\rm IR}$ decreases with increasing $z$ to stabilize to a roughly constant value at $z > 2$, consistent with the results of \citet{Algera2020}. The increase in $\bar{q}_{\rm IR}$ at $z\simgt 3.5$ has a low significance because of the limited statistics and the large uncertainties on photometric redshifts. However, we note that a rise in $\bar{q}_{\rm IR}$ at these redshifts can be expected due to the rapid growth  of the energy density of the cosmic microwave background (CMB),   $\epsilon_{\rm CMB}\propto (1+z)^4$  , with redshift. Correspondingly, the cooling time of relativistic electrons via inverse Compton scattering off the CMB rapidly decreases, leading to a dimming of synchrotron emission and therefore an increase in $\bar{q}_{\rm IR}$ at $z>3.5$ {\citep{Murphy2009, Lacki2010, SchleicherBeck2013}}.



\subsection{The \citet{Negrello2017} sample}

\citet{Negrello2017} selected from the \textit{Herschel} Astrophysical Terahertz Large Area Survey \citep[H-ATLAS;][]{Eales2010} a complete sample of 80 high-$z$ candidate strongly lensed  SFGs  hereafter referred to as the Negrello sample. {The sample includes all H-ATLAS galaxies with $500\,\mu$m flux density higher than 100\,mJy, except for local ones ($z\le 0.15$).}

Combining measurements by \citet{Negrello2017}, \citet{Neri2020}, \citet{Urquhart2022}, and \citet{Cox2023}, we collected spectroscopic redshifts for 60 of these galaxies. For the remaining {20} sources, we adopted the photometric redshifts estimated by \citet{Negrello2017}. {The redshifts range from 1.027 to 4.26 with a mean of 2.63. Values of the magnification derived for a small fraction of sources are in the range $\mu \sim 5$-15. The IR luminosities, demagnified by a typical $\mu\sim 10$, vary from $\log(L_{\rm IR}/L_\odot)= 12.1$ to 13.1 with a mean of 12.6; the SFRs range from $\sim 160$ to $\sim 1\,600\,M_\odot\,\hbox{yr}^{-1}$ with a mean of $\sim 510\,M_\odot\,\hbox{yr}^{-1}$.  For comparison, over the same redshift interval the Delhaize/MIGHTEE sample has $9.8 \le \log(L_{\rm IR}/L_\odot) \le 13.39$  , with a mean of $\log(\mu L_{\rm IR}/L_\odot)= 12.0$, and $0.19 \le \hbox{SFR}/M_\odot\,\hbox{yr}^{-1}\le 2\,360$. Thus, the IR luminosities and the SFRs of the Negrello sample are within the Delhaize/MIGHTEE range, albeit at the higher end of the range.  }

\citet{Giulietti2022} carried out 2.1\,GHz continuum observations with the Australia Telescope Compact Array (ATCA) of the 30 sources in the H-ATLAS South Galactic Pole field, with an angular resolution of $\sim 10''$. Of these, 11 were detected at $\ge 4\,\sigma$; their flux densities were scaled to 1.4\,GHz using a spectral index of $\alpha=-0.7$. For the other 19, we computed $4\,\sigma$ upper limits by measuring the rms noise at their positions on the 2.1\,GHz maps and scaling them to 1.4\,GHz with the mentioned spectral index and allowing a 10\% calibration uncertainty. 

The 50 candidate strongly lensed  SFGs in the H-ATLAS equatorial and North Galactic Pole (NGP) fields lie in the area covered by the FIRST survey. \citet{Giulietti2022} found a counterpart in the FIRST catalog for 8 of them; for 2 of these, more precise flux-density measurements were found in the literature and were adopted. Clear signals were detected in FIRST maps for 9 additional sources. We cross-matched the remaining 33 sources with the Rapid ASKAP Continuum Survey (RACS) catalog at 887.5\,MHz \citep{Hale2021} using a $5''$ search radius. 

Most of the RACS flux densities have a NON\_VALIDATED quality level flag (14 out of 18), with only four being flagged as UNCERTAIN. Rejecting these data would leave us with a low fraction of detections ($28/80=35\%$), making the determination of the mean/median $q_{\rm IR}$ highly uncertain. Therefore, we chose to accept as valid the RACS flux densities after having checked that there is reasonably good consistency between the RACS flux densities, scaled to 1.4\,GHz, and the FIRST flux densities for the 17 sources in the Negrello sample detected by both surveys. These are brighter on average, but have RACS quality levels similar to the 18 sources with only RACS detections. However, because of the limited angular resolution of the RACS survey ($25''$), the flux density may be contaminated by nearby sources. For this reason, we dropped the three matches whose flux densities ---scaled to 1.4 GHz assuming a spectral index of $-0.7$--- exceed the FIRST completeness limit of 1\,mJy ($5''$ resolution). For the remaining 18 NGP plus equatorial sources, we adopted 1\,mJy as the upper limit to $S_{1.4\,\rm GHz}$. 
In total, we have 43 detections ($53.75\%$): 28 from \citet{Giulietti2022} 
and 15 from the RACS catalog. 

There are only four Negrello sources at $z>4$, three of which are undetected in the radio. Therefore, we considered only the 76 sources at $z<4$. These were split into four redshift bins. The results of the ASURV analysis for each bin are reported in Table\,\ref{tab:Negrello_qIR}. We find no significant redshift dependence of the mean or median $q_{\rm IR}$ . The derived mean and median values are significantly higher than those found for the Delhaize/MIGHTEE sample and by \citet[][see below]{Algera2020}.

{Somewhat higher values of $\bar{q}_{\rm IR}$ for the Negrello sample are expected as a consequence of its submm selection (which leads to a selection bias for IR-bright galaxies), while the Delhaize/MIGHTEE sample combines radio and IR selections. Differential magnification can further} bias high the values of $q_{\rm IR}$ because submm surveys show a strongly selection bias for caustics close to compact star-formation regions, boosting their flux densities \citep{Serjeant2012, Hezaveh2012}. As illustrated by Fig.\,8 of \citet{Lapi2012}, the maximum amplification for an extended source depends on its size, being higher for more compact sources. In the likely case where the size of the radio emission is substantially larger than the star-formation region \citep[cf., e.g.,][]{Thomson2019}, the amplification of the latter is correspondingly larger.  Also, the IR emission may be closer to the optical axis than the radio emission; this yields a further increase in the relative magnification.




\subsection{The \citet{Algera2020} sample}

\citet{Algera2020} determined the median $q_{\rm IR}$ in 15 redshift bins in the range 1.5--4 for a sample of 647 SMGs (12 radio-excess AGN excluded) drawn from  Atacama Large Millimeter Array (ALMA) SCUBA-2 Ultra Deep Survey {at $870\,\mu$m \citep[AS2UDS;][]{Stach2019} with $S_{870\,\mu\rm m}> 0.9\,$mJy}. The field was covered by deep 1.4\,GHz Karl G. Jansky VLA observations (rms noise in the range 7--$10\,\mu\hbox{Jy}\,\hbox{beam}^{-1}$) with an angular resolution of $1''.8 \times 1''.6$. 
Radio-undetected sources were taken into account by stacking together detected and undetected SMGs and using the median values. The IR luminosity was derived by MAGPHYS SED fitting.

{The UV to radio SEDs were determined by \citet{Dudzeviciute2020} using the MAGPHYS code, exploiting deep optical/near-IR ground-based photometry and \textit{Spitzer} IRAC and MIPS observations in addition to ALMA measurements. Moreover, \textit{Herschel} PACS and SPIRE maps were deblended adopting ALMA, \textit{Spitzer} $24\,\mu$m, and 1.4 GHz astrometry  as
positional priors.  The IR luminosities cover the range $\log(L_{\rm IR}/L_\odot)\simeq 11.8$--13.3, with a median of 12.46, which is close to that of the Negrello sample. The SFRs span the interval from $\simeq 40$ to $\simeq 1,000\,M_\odot\,\hbox{yr}^{-1}$ with a median of $290\,M_\odot\,\hbox{yr}^{-1}$ \citep{Dudzeviciute2020}.}

No significant variation of the median $q_{\rm IR}$ with redshift was found. The mean for the full sample is $\bar{q}_{\rm IR}=2.20\pm 0.03$. A similar result was found for a luminosity-limited subsample, with $L_{\rm IR}$ in the range 4--$7\times 10^{12}\,L_\odot$ and at  least one \textit{Herschel}/SPIRE detection, granting more solid constraints on $L_{\rm IR}$.






\section{Discussion and conclusions}\label{sect:conclusions}

\begin{figure}
        \includegraphics[width=\linewidth]{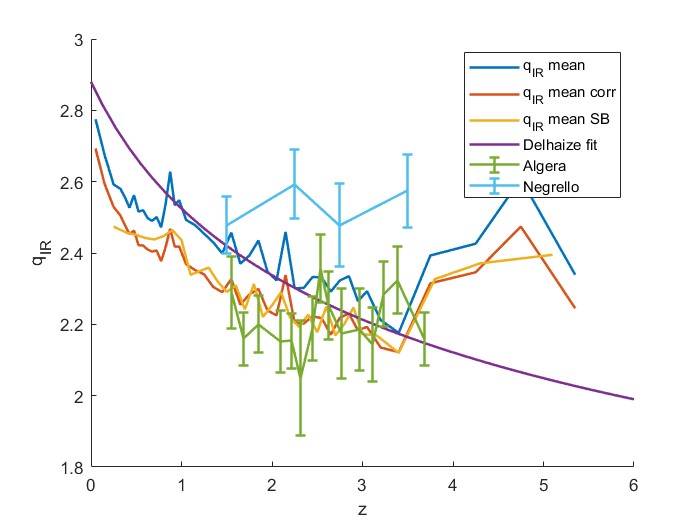}
  \caption{Redshift dependence of $\bar{q}_{\rm IR}$. The ``$q_{\rm IR}$ mean'',  ``$q_{\rm IR}$ mean corr'', and ``$q_{\rm IR}$ mean SB'' lines refer, respectively, to the initial Delhaize/MIGHTEE sample, the same sample but with $S_{1.4\,\rm GHz}$ flux densities and upper limits derived from the 3\,GHz survey scaled up by a factor of 1.51, and to this latter sample after removing SMGs with $\hbox{SFR}<10\,M_\odot\,\hbox{yr}^{-1}$ (``starburst plus protospheroid'' sample). Error bars are omitted so as to avoid overcrowding. The ``Delhaize fit'' line shows the power-law fit, $\bar{q}_{\rm IR}(z)=2.88(1+z)^{-0.19}$, obtained by \citet{Delhaize2017}. The results of our analysis of the \citet{Negrello2017} sample and of the study of \citet{Algera2020} are also shown.}
  \label{fig:qIRvs_z}
\end{figure}

\begin{figure}
        \includegraphics[width=\linewidth]{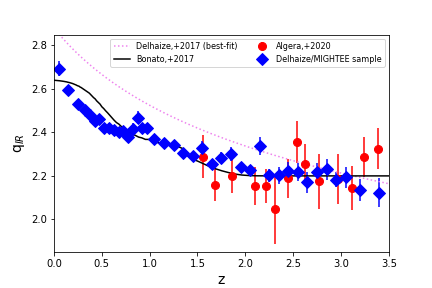}
  \caption{Comparison of the redshift dependence of $\bar{q}_{\rm IR}$ derived from the \citet{Bonato2017} model (solid black line) with determinations from the Delhaize/MIGHTEE sample and with the results by \citet{Algera2020}. The error bars for the Delhaize/MIGHTEE sample, which are in many cases lower than the symbol size, are the Poisson uncertainties provided by the ASURV package. The true errors, including contributions from uncertainties on photometric redshifts and from the incomplete data used to estimate IR and radio luminosities, are far larger but difficult to quantify. The model values of $\bar{q}_{\rm IR}$ were computed adopting mean values of $q_{\rm IR}$ of 2.64, 2.35, and 2.2 for normal late-type, starburst, and protospheroidal galaxies, respectively, and weighting with the redshift-dependent abundances shown in Fig.\,\protect\ref{fig:zdistr}.
}
  \label{fig:qIRmodel}
\end{figure}

\begin{figure}
        \includegraphics[width=\linewidth]{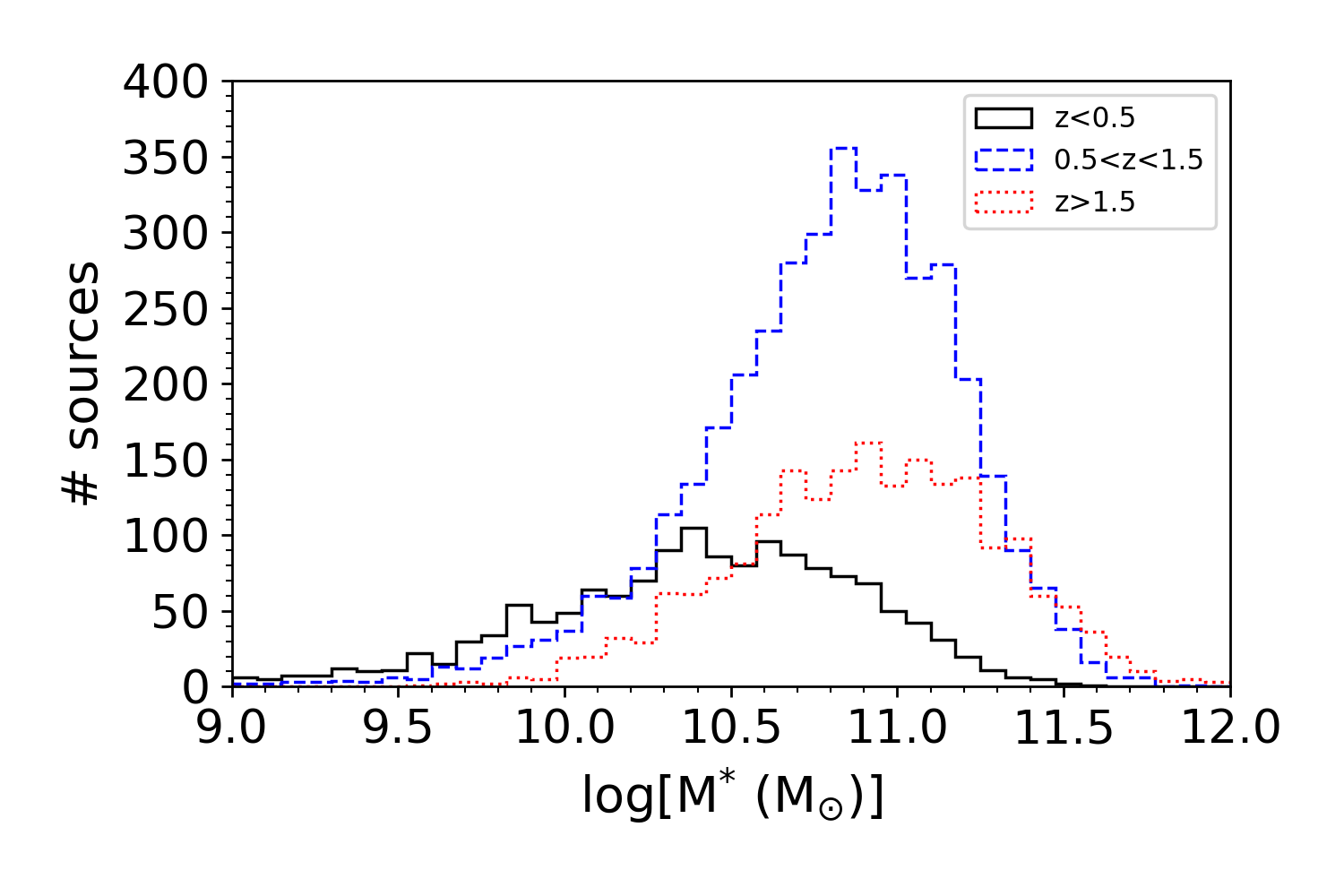}
  \caption{Distribution of stellar masses of SFGs in the Delhaize/MIGHTEE sample in three redshift intervals ($z<0.5$, $0.5<z<1.5$, and $z>1.5$), dominated, according to our reference model, by normal disk, starburst, and protospheroidal galaxies, respectively.}
  \label{fig:Mstar}
\end{figure}

Figure\,\ref{fig:qIRvs_z} summarizes our results regarding the redshift dependence of $\bar{q}_{\rm IR}$ obtained from the analyses presented in the previous section. The large radio- and IR-selected sample by \citet{Delhaize2017} is central to this study, as it covers a broad redshift range. We have updated the analysis by taking into account the  MIGHTEE survey of most of the COSMOS field \citep{Heywood2022}, which has allowed us to increase the fraction of radio-detected, no-radio-excess SFGs in this area from 57.3\% to 66.7\%, thus improving the accuracy of the determination of $\bar{q}_{\rm IR}(z)$.

We detect an offset between the 1.4\,GHz flux densities inferred from the 3\,GHz and those from the MIGHTEE surveys. After accounting for this, the mean $q_{\rm IR}$ of local SFGs decreases from $2.88\pm 0.03$, as found by \citet{Delhaize2017}, to $2.693\pm 0.035$. The median value is 2.646, in good agreement with earlier determinations \citep[cf., e.g.,][who found a median value of $2.64\pm 0.02$]{Bell2003}.

The mean $q_{\rm IR}$ steadily decreases  with increasing $z$ up to $z\simeq 2$. At higher redshifts $\bar{q}_{\rm IR}$ stabilizes at $\sim 2.2$, in good agreement with the findings by \citet{Algera2020}. At these redshifts, spheroidal galaxies rapidly form most of their stars at high rates.

Our analysis of the Negrello sample of strongly lensed galaxies selected at $500\,\mu$m yields values of $\bar{q}_{\rm IR}$ that are higher than those derived from the Delhaize/MIGHTEE sample and found by \citet{Algera2020} {in the same redshift range}. The significance of the discrepancy is hard to assess because the fraction of Negrello sources with accurate radio measurements is relatively low. However, we argue that the difference could be accounted for by differential magnification, {suggesting that caution must be exercised  when exploiting strongly lensed data for this kind of analysis}.

Figure\,\ref{fig:qIRmodel} compares the observational determinations with the redshift dependence of $\bar{q}_{\rm IR}$ for the mixture of normal late-type, starburst, and protospheroidal galaxies yielded by the \citet{Bonato2017} model for $S_{1.4\,\rm GHz} > 10\,\mu$Jy, which is approximately five times the rms fluctuations of the MIGHTEE survey of the COSMOS field \citep{Heywood2022}. The three galaxy populations were attributed redshift-independent values. {For normal galaxies, we adopted the local value obtained by \citet{Bell2003}, $\bar{q}_{\rm IR}= 2.64$; for protospheroids, we adopted $\bar{q}_{\rm IR}= 2.2$ following \citet{Algera2020}; and for starbursts, we adopted  $\bar{q}_{\rm IR}= 2.35$ based on values obtained around $z\simeq 1$, where this population dominates according to the \citet{Bonato2017} model (see Fig.\,\ref{fig:zdistr}). }

\citet{Bonato2017} presented predictions for the evolution of the radio luminosity function based on the model by \citet{Cai2013}. The latter model deals with the co-evolution of galaxies and active galactic nuclei (AGN) from optical to millimeter wavelengths. A phenomenological approach was used for normal and starburst galaxies, while a physical model was developed for ``spheroidal'' galaxies. \citet{Bonato2017} converted the redshift-dependent SFR functions computed by \citet{Cai2013} into SFG radio luminosity functions. These authors used a constant ratio between SFR and  free-free emission (their eq.\,(1)) and different recipes for the relation between SFR and synchrotron luminosity. We selected the model using the mildly nonlinear relationship ($L_{\rm sync}\propto \hbox{SFR}^{1.1}$, their eq.\,(5)) with a dispersion of $\log(L_{\rm sync})$ at fixed SFR $\sigma_{\log(\rm Lsync)} = 0.3$.

{The distribution of the log of 1.4\,GHz flux densities for the Delhaize/MIGHTEE sample peaks at $47\,\mu$Jy and drops rapidly at fainter flux densities, implying significant incompleteness. As most sources (66.7\%) are already detected, we can safely guess that (almost) all of them should be detected by a survey complete down to $10\,\mu$Jy. A similar conclusion holds for the other samples. In other words, the model predictions also include (almost) all undetected sources, which were taken into account in the data analysis by means of a survival analysis. }

The agreement between the model and the observational estimate of $\bar{q}_{\rm IR}(z)$ is relatively good, confirming that the variation with redshift of the galaxy population dominating the cosmic star-formation activity using different morphologies and, presumably, different configurations of the magnetic field can provide a simple explanation for the redshift dependence of $\bar{q}_{\rm IR}$.
No significant weight should be given to the excess of the model over the observational estimate at $z\simlt 0.6$ and to the increase in $\bar{q}_{\rm IR}$ at $z\simgt 4$. As mentioned above, the error bars are purely statistical and do not include the substantial uncertainties on photometric redshifts and on estimates of IR and radio luminosities based on incomplete data. However, if confirmed, the low-$z$ excess may suggest an additional parameter is needed, such as the stellar mass.

The relatively high values of $\bar{q}_{\rm IR}$ at high $z$ may suggest the onset of radio dimming due to increasing inverse Compton losses of relativistic electrons off the CMB, whose energy density is proportional to $(1+z)^4$. The onset redshift is difficult to predict due to the complexity of involved processes and to the poor knowledge of the physical parameters describing the interstellar medium and the magnetic fields. However, the values we find are within the range of values yielded by models \citep{Murphy2009, SchleicherBeck2013, Schober2016, Schober2023}.

We note that this interpretation is consistent with evidence of higher radio-to-IR-luminosity ratios (lower values of $\bar{q}_{\rm IR}$) for more massive galaxies \citep{Gurkan2018, Delvecchio2021, Smith2021, McCheyne2022}: as mentioned in Sect.\,\ref{sect:introduction}, spheroidal galaxies, which dominate the star formation at high $z$, have {higher stellar velocity dispersions and therefore higher stellar masses}.

Figure\,\ref{fig:Mstar} shows the distributions of stellar mass, $M_\star$, {in the Delhaize/MIGHTEE sample} in the redshift ranges of $z<0.5$, $0.5<z<1.5$, and $z>1.5$, which ---according to our reference model---  are dominated by {normal late-type galaxies, starbursts, and protospheroids}, respectively (cf. Fig.\,\ref{fig:zdistr}). For $\log(M_\star/M_\odot),$ we have used the values labeled ``mstar\_best'' in the catalog by \citet{Delhaize2017}. The mean values of $\log(M_\star/M_\odot)$ are 10.39, 10.79, and 10.90, respectively. The modest variations of mean values imply that stellar-mass dependencies of $q_{\rm IR}$ reported in the literature are unlikely to fully  account for the observed variations of $q_{\rm IR}$ among the populations, {leaving room for our proposal} that such variations {may be} related to their  different structural properties.


\begin{acknowledgements}
{We are grateful to the referee for an in-depth review of the paper and useful comments.}
MB acknowledges support from the INAF mini grant ``A systematic search for ultra-bright high-z strongly lensed galaxies in Planck catalogues''.
HSBA was supported by NAOJ ALMA Scientific Research Grant Code 2021-19A (HSBA). This research has made use of the TOPCAT tool \citep[\url{https://www.star.bristol.ac.uk/mbt/topcat/};][]{Taylor2005}.
\end{acknowledgements}

\section*{Data Availability}

The data used in this paper are available in the referenced literature. The derived data generated in this research will be shared on request to the first author.

\bibliographystyle{aa} 
\bibliography{FIRRC}

\end{document}